\def\Ar{\rightarrow}

\def\a{\alpha}
\def\b{\beta}

\def\m{\mu}

\def\th{\theta}

\def\l{\lambda}

\def\eV{{\rm eV}}
\documentstyle [seceq,epsf,preprint]{ptptex}
\def\dfrac{\displaystyle\frac}
\begin{document}
\renewcommand{\thefootnote}{\fnsymbol{footnote}}
\baselineskip=24pt
\setcounter{page}{1}
\thispagestyle{empty}
%\topskip 1  cm
%\topskip 0.5  cm
%\begin{flushright}
%\begin{tabular}{c r}
%& hep-ph/0006099\\
%& KEK-TH-700, NIIG-DP-00-04 \\
%& June 2000, Dec.2000(Rev.)
%\end{tabular}
%\end{flushright}
\markboth{Naotoshi Okamura, Morimitsu Tanimoto}
{Generic Relations of Flavor Mixings between Leptons 
and Quarks in SU(5)}
%\vspace{1 cm}
%\centerline{\LARGE \bf Generic Relations of Flavor Mixings}
%\vskip 0.5 cm
%\centerline{\LARGE\bf  between Leptons  and Quarks  in SU(5)}
\title{
Generic Relations of Flavor Mixings
\\
between Leptons  and Quarks  in SU(5)
}
%\vskip 1.2 cm
\author{
Naotoshi Okamura$^A$
  \footnote{E-mail address: naotoshi.okamura@kek.jp}
\\
and
\\
Morimitsu Tanimoto$^B$
  \footnote{E-mail address: tanimoto@muse.hep.sc.niigata-u.ac.jp}
}
%\vskip 0.2 cm
\inst{$^A$Theory Group, KEK, Tsukuba, 
 Ibaraki 305-0801950-2181, JAPAN
\\
$^B$Department of Physics, Niigata University, 
 Niigata 950-2181, JAPAN
}
%\vskip 0.3 cm
%\centerline{\large \bf  and}
%\vskip 0.3 cm
%
%\vskip 1.5 cm
%\centerline{\bf ABSTRACT}\par
%\vskip 0.2 cm
\abst{
 We have studied  implications of the generic
lopsided mass matrix of the charged leptons by taking the SU(5) GUT relation
  in the nearest-neighbor interaction (NNI) basis.
  We have found four  interesting relations among  the lepton mixings
and the quark ones, which are independent of details of the model.
These  relations  are discussed by using the experimental data.
  We have also discussed the relation between  $U_{e2}$ and $U_{e3}$
 incuding the contribution from the neutrino mass matrix.  
We have presented  the probable value
  $U_{e3}=0.05\sim 0.16$, which is  
independent of the solar neutrino solutions. 
 The CP violating quantity $J_{CP}$ is also discussed.
}
\preprintnumber[35ex]{
hep-ph/0006099\\
KEK-TH-700, NIIG-DP-00-04 \\
June 2000, Sept.2000(Rev.)
}
\notypesetlogo
\maketitle
%\newpage
%%%%%%%%%%%%%%%%%%%%%%%%%%%%%%%%%%%%%%%%%%%%%%%%%%%%%%%%%%%%%%%%%%%%%%%%%%%%%%%
\renewcommand{\thefootnote}{\fnsymbol{footnote}}
%%%%%%%%%%%%%%%%%%%%%%%%%%%%%%%%%%%%%%%%%%%%%%%%%%%%%%%%%%%%%%%%%%%%%%%%%%%%%%%
\topskip 0. cm
\section{Introduction}

  Recent Super-Kamiokande data of atmospheric neutrinos \cite{SKam} have provided  a more solid evidence of the neutrino oscillation, which corresponds to 
  the nearly maximal flavor mixing of the neutrinos.
 The large  mixing  gives the strong  constraint on the structure of the 
 lepton mass matrices in the three family model.
 The flavor mixing angles of the neutrinos are mismatch between the
eigenstates of the neutrinos and those of the charged leptons, in other
words between the mass matrices of the neutrinos and the charged
leptons.
One interesting idea is that the charged lepton mass matrix ($M_L$) is related
with the down quark one ($M_D$) such as $M_L=M_D^T$,
which is given by the unified  or the flavor symmetry 
\cite{lopside,OH,Albright,BKY}.
Then, the large right-handed mixing for the quarks leads to the large
left-handed mixing for the leptons. The large left-handed  lepton
mixing with leaving the small left-handed quark mixing leads to
the ``lopsided'' structure  of the mass matrices \cite{Barr}.
The model is expected to present  relations among  the lepton mixings and 
the quark ones.
Actually,  a few  authors have found interesting  relations 
in $SO(10)$ and $E_6$ \cite{Albright,BKY}.

In this paper, we study  implications of the generic lopsided mass
matrix by taking the  nearest-neighbor interaction (NNI) basis
\cite{NNI}. In this basis, one can take the lopsided texture for 
the down-quarks without of loss generality. 
Assuming the nontrivial relation $M_L=M_D^T$ of the SU(5) GUT
in the NNI basis, 
we examine relations among the lepton mixings and the quark ones.
If the neutrino mass matrix is close to the diagonal one,
in other words, the mixings from the neutrino sector 
are  negligibly small, 
we find  some  relations among  the lepton mixings and the quark ones,
which are  independent of  details of the model.
These relations will be tested precisely in the near future.
The deviations of  these relations are also discussed
taking account  the effect of the non-diagonal neutrino mass matrix.
Then, the magnitude of the CP violation in the lepton sector is also
discussed.

In section 2, we discuss the quark mass matrices in the NNI  basis.
In section 3, the relations  between  the flavor mixings of the quarks
and the leptons are derived in the SU(5) GUT.
In section 4, discussions and summary are given.

%%%%%%%%%%%%%%%%%%%%%%%%%%%%%%%%%%%%%%%%%
%%%%%%%%%%%%%%%%%%%%%%%%%%%%%%%%%%%%%%%%%
\section{Quark Mass Matrices in the NNI basis}

As presented by Branco, Lavoura and Mota \cite{NNI}, 
both up- and down-quark mass
matrices could always be transformed to the non-Hermitian matrices
in the NNI basis for three families.
 Takasugi \cite{Taka} has shown that quark mass matrices can be transformed
in general to the Fritzsch type parameterization \cite{Fritzsch} 
for the up-quarks  with retaining the NNI form for the down-quarks.
  In this basis, several authors have found that  
the down-quark mass matrix has a lopsided structure by studying the quark
masses and Cabibbo-Kobayashi-Maskawa (CKM) matrix \cite{CKM} 
phenomenologically \cite{Okamura,Ito}.
%%%%%%%%%%%%%%%%%%%%%%%%%%%%%%%%%%%%%%%

   This basis is consistent with the $SU(5)$ GUT because the up-quark mass
matrix is the symmetric one.  In the  $SU(5)$ GUT with 
$\bf 5$ and $\bf 5^*$ Higgs multiplets, the charged lepton mass
matrix  $M_L$ is related to the down quark one $ M_D$ at the GUT scale
as follows:
\begin{equation}
      M_L=M_D^T \ ,
 \label{GUT}
\end{equation}
\noindent
which gives the nearly maximal mixing of the left-handed unitary
matrix for the charged lepton.  On the other hand,
the neutrino mass matrix is independent of  other fermion matrices since
the neutrino mass is  given as $\bf 5^* 5^*$ in the  $SU(5)$ GUT
\footnote{The neutrinos are supposed to be Majorana particles.}.
Thus, the lepton flavor mixing, so called the
Maki-Nakagawa-Sakata (MNS) mixing matrix
$V_{\a i}$ \cite{MNS},
is undetermined due to lack of information of the neutrino
sector in spite of  eq.(\ref{GUT}). 

This situation is understandable
 because we only choose a convenient basis for quark sectors
 in order to give the maximal mixing of the left-handed charged lepton
 through eq.(\ref{GUT}). 
There is a lot of freedoms in the neutrino mass matrix.
 Once a  model as to the  flavor structure is put, the neutrino mass
matrix is fixed,  and then, one can predict the MNS mixings.
 However, almost  models with the lopsided structure of the charged
lepton mass matrix \cite{lopside} give small  mixings in the neutrino
mass matrix
%%%%%%%%%%%%%%%%%%%%%%%%%%%%%%%%%%%%%%%%%%%%%%%%%%%%%%%%
\footnote{Nomura and Yanagida \cite{lopside} built a model
with the maximal mixing between the first and second family in the
neutrino mixing matrix. This case corresponds to the bi-maximal mixings.}.
%%%%%%%%%%%%%%%%%%%%%%%%%%%%%%%%%%%%%%%%%%%%%%%%%%%%%%%%
 In these models, the MNS matrix
is mainly determined by the charged lepton mass matrix.
Therefore, the NNI basis in eq.(\ref{GUT}), which was  taken by
Hagiwara and Okamura  \cite{OH}, seems to be a  physically meaningful one .

We work in the  NNI basis with eq.(\ref{GUT}), and so we investigate  the
 relations between the CKM and the MNS mixings neglecting the
contribution from the neutrino mass matrix as the first step.
Generic relations are also discussed 
including the contribution of the neutrino sector.

%%%%%%%%%%%%%%%%%%%%%%%%%%%%%%%%%%%%%%%%
%%%%%%%%%%%%%%%%%%%%%%%%%%%%%%%%%%%%%%%%%

 We assume that oscillations need only account for 
 the solar and the atmospheric neutrino data.
  Our starting point as to the neutrino mixing is
  the large $\nu_\mu \Ar \nu_\tau$ oscillation of  the 
atmospheric neutrinos  with 
 $\Delta m^2_{\rm atm}=  (2\sim 5)\times  10^{-3} \eV^2$ and 
 $\sin^2 2\th_{\rm atm} \geq 0.88$,
 which is derived from the recent data of the atmospheric neutrino deficit 
 at Super-Kamiokande \cite{SKam}. 
 For the solar neutrinos \cite{SKamsolar}, 
 we take into account of  still allowed   solutions,
   the small mixing angle (SMA) MSW \cite{MSW}, 
the large mixing angle (LMA) MSW, the low $\Delta m^2$ (LOW)  and  
the vacuum oscillation (VO) solutions \cite{BKS}.
	
%%%%%%%%%%%%%%%%%%%%%%%%%% Up Quark %%%%%%%%%%%%%%%%%%%%

We begin with writting  the quark mass matrices in the NNI-form without
loss of generality.  Following ref.\cite{Taka}, the up-quark  mass
matrix ia taken as  the Fritzsch
texture \cite{Fritzsch} at the SU(5) GUT scale:
\begin{equation}
M_{\rm U}^{} = m_3^{}
\left(
\begin{array}{ccc}
        0 & a^{}_{\rm u} & 0 \\
a^{}_{\rm u} &       0   & b^{}_{\rm u} \\
        0 & b^{}_{\rm u} & c^{}_{\rm u}
\end{array}
\right)\,,
\label{up-mass}
\end{equation}
where $m_3^{}$ is the third family  mass and
each parameter ($a^{}_{\rm u},b^{}_{\rm u},c^{}_{\rm u}$) is  
written as
\begin{eqnarray}
a^{}_{\rm u} &=& \sqrt{
              \dfrac{m_1^{} m_2^{}}
                    {m_3^{}\left(m_3^{}-m_2^{}+m_1^{}\right)}} 
          \simeq \dfrac{\sqrt{m_1^{} m_2^{}}}{m_3^{}}\,, \nonumber \\
b^{}_{\rm u} &=& \dfrac{1}{m_3^{}}
         \sqrt{\dfrac{\left(m_3^{} - m_2^{} \right)
                      \left(m_3^{} + m_1^{} \right)
                      \left(m_2^{} - m_1^{} \right)}
	             {\left(m_3^{} - m_2^{} + m_1^{} \right)}} 
         \simeq\sqrt{\dfrac{m_2^{}}{m_3^{}}}\,, \nonumber \\
c^{}_{\rm u} &=& 1-\dfrac{m_2^{}}{m_3^{}}+\dfrac{m_1^{}}{m_3^{}} 
          \simeq 1\,,
\label{elements_up}
\end{eqnarray}
where $m_2^{}$ and $m_1^{}$ stand for the second and the first family
up-type quark masses, respectively.
All elements in the up-quark mass matrix can be taken real numbers 
without losing generality in this basis, and 
$a_u \ll b_u \ll c_u$ is obtained due to the up-quark mass hierarchy.

%%%%%%%%%%%%%%%%%%%%%%%%%%%%%%%%%%%%%%%%%%%%%%%%%%%%%%%%%%%%%%%%%%%
%%%%%%%%%%%%%%%%  Down Quarks %%%%%%%%%%%%%%%%%%%%%%%%%%%%%%%%%%%%%

On the other hand, down-quark mass matrix, which is
the NNI-form texture  at the SU(5) GUT scale, is
\begin{equation}
M_{\rm D}^{} = m_3^{}
\left(
\begin{array}{ccc}
        0 & a^{}_{\rm d} e^{i\theta_{\alpha}}& 0 \\
c^{}_{\rm d}e^{-i\theta_{\alpha}} &    0   & b^{}_{\rm d}e^{i\theta_{\beta}} \\
        0 & d^{}_{\rm d} & e^{}_{\rm d}
\end{array}
\right)\,,
\label{down-mass}
\end{equation}
where $m_3^{}$ is the third family mass and all parameters are  real.
The phase assignment in eq.(\ref{down-mass}) is generic one.
In terms of  down-quark  mass ($m_3^{}$, $m_2^{}$, $m_1^{}$)
and two dimension-less parameters ($y^{}_{\rm d}$, $z^{}_{\rm d}$),
 which is assumed to be of order one,      
all parameters in eq.(\ref{down-mass}) are written as \cite{Okamura}
\begin{eqnarray}
 a^{}_{\rm d} &=& q^{}_{\rm d} \dfrac{z^{}_{\rm d}}{y^{}_{\rm d}} 
               = \dfrac{\sqrt{m_2^{} m_1^{}}}{m_3^{}}
                    \dfrac{z^{}_{\rm d}}{y^{}_{\rm d}}\,, \nonumber \\
 b^{}_{\rm d} &=& \sqrt{\dfrac{p_d}{(1-y_{\rm d}^4)}} 
           \simeq \dfrac{m_2^{}}{m_3^{}} \dfrac{1}{\sqrt{1-y_{\rm d}^4}}\ ,
       \nonumber \\
 c^{}_{\rm d} &=& q^{}_{\rm d} \dfrac{1}{z^{}_{\rm d} y^{}_{\rm d}} 
               = \dfrac{\sqrt{m_2^{} m_1^{}}}{m_3^{}}
                    \dfrac{1}{z^{}_{\rm d}y^{}_{\rm d}}\,, \nonumber \\
 d^{}_{\rm d} &=&\sqrt{1-y_{\rm d}^4 }\,,\nonumber \\
 e^{}_{\rm d} &=& y^2_{\rm d} \ ,
\label{elements_down}
\end{eqnarray}
where $p^{}_{\rm d}$ and $q^{}_{\rm d}$ are defined as
\begin{eqnarray}
 p^{}_{\rm d} &=& \dfrac{m_1^2 + m_2^2}{m_3^2} \ , \nonumber \\
 q^{}_{\rm d} &=& \dfrac{\sqrt{m_1^{} m_2^{}}}{m_3^{}} \ .
\end{eqnarray}
\noindent
Thus, five unknown parameters are given by three down quark masses
and the parameters $(y_d,\  z_d)$.

%%%%%%%%%%%%%%%%%%%%%%%%%%%%%%%%%%%%%%%%%%%%%%%%%%%%%%%%%%%%%
%%%%%%%%%%%%%%%%% CKM elements %%%%%%%%%%%%%%%%%%%%%%%%%%%%%

By taking   $M_{\rm U}M_{\rm U}^{\dagger}$ and 
$M_{\rm D}M_{\rm D}^{\dagger}$,  one obtains 
the CKM matrix elements  up to $O(\l^2)$,  where $\l$
is the Cabibbo angle, in terms of quark masses and parameters
 $y^{}_{\rm d}$, $z^{}_{\rm d}$, $\theta$:
\begin{eqnarray}
%\left|
V_{us}
%\right|
 &\simeq&  
%\left|
 y^{}_{\rm d} z^{}_{\rm d}\sqrt{\dfrac{m_d}{m_s}}
- \sqrt{\dfrac{m_u}{m_c}}e^{i\theta} \ ,
%\right|\ ,
\label{Vus}\\
%
%\left|
V_{cb}
%\right|
 &\simeq&  
%\left|
\dfrac{y^2_{\rm d}}{\sqrt{1-y^4_{\rm d}}}
\dfrac{m_s}{m_b}
e^{i\theta}
-
\sqrt{\dfrac{m_c}{m_t}}
e^{-i\theta_\alpha}
%\right|
\,,
\label{Vcb} \\
%
%%\left|
V_{ub}
%\right|
 &\simeq&  
%\left|
y^{}_{\rm d}z^{}_{\rm d}
\dfrac{\sqrt{1-y^4_{\rm d}}}{y^2_{\rm d}}
\sqrt{\dfrac{m_d}{m_s}}\dfrac{m_s}{m_b}
+
\sqrt{\dfrac{m_u}{m_c}}V_{cb} \ ,
%\right|
%\left|
%\dfrac{z^2_{\rm d}\sqrt{1-y^4_{\rm d}}}{y^{}_{\rm d} z^{}_{\rm d}}
%\sqrt{\dfrac{m_d}{m_s}}\dfrac{m_s}{m_b}
%+
%\sqrt{\dfrac{m_u}{m_c}}
%\dfrac{y^2_{\rm d}}{\sqrt{1-y^4_{\rm d}}}
%\dfrac{m_s}{m_b}
%e^{i\theta}
%\right|\,,
\label{Vub}
% \\
%
%\left|\dfrac{V_{td}}{V_{ts}}\right| &\simeq&  
%\dfrac{z^{}_{\rm d} y^{}_{\rm d}}{y^4_{\rm d}}
%\sqrt{\dfrac{m_d}{m_s}}
%\label{VtdVts}\,,
\end{eqnarray}
where $\theta$ is defined as
\begin{equation}
 \theta=\theta_{\beta}-\theta_{\alpha}\ .
\end{equation}
\noindent
 Thus, the CKM matrix elements are given in terms of mass eigenvalues
and four parameters $y_d$, $z_d$, $\theta_\a$ and $\theta_\b$.
These equations are satisfied up to $O(\lambda^2)$.
It is emphasized that $y_d^4\simeq 1/2$ leads to  the lopsided down quark
mass matrix, which is consistent with the experimental data.

%%%%%%%%%%%%%%%%%%%%%%%%%%%%%%%%%%%%%%%%%%%%%%%%%%%%%%%%%%%%
%%%%%%%%%%%%%  Lepton %%%%%%%%%%%%%%%%%%%%%%%%%%%%%%%%%%%%%%
%%%%%%%%%%%%%%%%%%%%%%%%%%%%%%%%%%%%%%%%%%%%%%%%%%%%%%%%%%%
\section{Relations of Flavor Mixings in SU(5)}

In this basis, we assume the SU(5) GUT relation
between the down-quark mass matrix and the charged lepton one.
Taking account of  eq.(\ref{GUT}), 
$M_{\rm L}^{}M_{\rm L}^{\dagger}$ is written as
\begin{equation}
 M_{\rm L}^{}M_{\rm L}^{\dagger}
=M_{\rm D}^{T}M_{\rm D}^{\ast}
=
\left(
\begin{array}{ccc}
c^2_{\rm d} & 0 & c^{}_{\rm d}b^{}_{\rm d}
                      e^{-i\left(\theta_{\alpha}+\theta_{\beta}\right)} \\
          0 & a_{\rm d}^2 + d_{\rm d}^2 & d^{}_{\rm d}e^{}_{\rm d} \\
c^{}_{\rm d}b^{}_{\rm d}e^{i\left(\theta_{\alpha}+\theta_{\beta}\right)} &
 d^{}_{\rm d}e^{}_{\rm d}  & b_{\rm d}^2 + e_{\rm d}^2
\end{array}
\right)\ ,
\label{chargedlepton-mass}
\end{equation}
which  is diagonalized by the following unitary matrix $U^{\rm e}$:
%%%%%%%%%%%%%%%%%%%%%%%%%%%%%%%%%%%%%%%%%%%%%%%%%%%%%

\begin{equation}
\left(U^{\rm e}\right)^{\dagger}=
\left(
\begin{array}{ccc}
1 & -\sqrt{\dfrac{m_1^{}}{m_2^{}}}\dfrac{y^2}{yz}
  &  \sqrt{\dfrac{m_1^{}}{m_2^{}}}\dfrac{\sqrt{1-y^4}}{yz} \\
-\dfrac{1}{yz}\sqrt{\dfrac{m_1^{}}{m_2^{}}}\dfrac{m_2^{}}{m_3^{}}
  &  y^2 & -\sqrt{1-y^4} \\
\dfrac{y}{z\sqrt{1-y^4}}\dfrac{m_2^{}\sqrt{m_1^{}m_2^{}}}{m_3^2}
  &  \sqrt{1-y^4} & y^2
\end{array}
\right)
\left(
\begin{array}{ccc}
e^{i\left(\theta_{\alpha}+\theta_\beta\right)} & 0 & 0 \\
\medskip \\
0& 1& 0\\
\medskip \\
0& 0& 1
\end{array}
\right) + {O}(\lambda^2)
 \label{unitary_e}
\end{equation}
where $y = y_{\rm d}^{}$, $z=z_{\rm d}^{}$ and $m_i^{}$ is the 
$i$-th family  charged-lepton masses at the SU(5) GUT scale.
%%%%%%%%%%%%%%%%%%%%%%%%%%%%%%%%%%%%%%%%%%%%%%%%%%%%%%
Due to  $M_{\rm e}^{}=M_{\rm d}^T$, the charged-lepton masses
are same as the down-quark masses. In order to get the realistic 
charged-lepton
mass hierarchy, the higher dimensional mass operator may be added
in the mass terms.  We  take account  this effect 
 by using  the physical charged lepton masses since $y$ and $z$
are independent of the mass eigenvalues. 
Then, $y$ and $z$ in the leptons are any more  different from the ones in the 
down-quarks.
%%%%%%%%%%%%%%%%%%%%%%%%%%%%%%%%%%%%%%%%%%%%%%%%%%%%%%%%%%%%
With the expresion  $U^{\rm e}_{\a i}$ ($\a=e,\m,\tau$; $i=1,2,3$)
for the unitary matrix  $(U^{\rm e})^{\dagger}$,
from  eqs.(\ref{Vcb}) and  (\ref{unitary_e})  we obtain 
two relations:
\begin{equation}
\cot \theta_{\mu\tau}^e \equiv
\left|\dfrac{U^{\rm e}_{\tau 3}}{U^{\rm e}_{\mu 3}}\right|
=
\left|\dfrac{U^{\rm e}_{\mu 2}}{U^{\rm e}_{\mu 3}}\right|
=
\left|\dfrac{U^{\rm e}_{e 2}}{U^{\rm e}_{e 3}}\right|
\label{relation1}\,,
\end{equation}
\noindent and 
\begin{equation}
 \cot \theta_{\mu\tau}^e =
%\dfrac{m_b}{m_s}\left|V_{cb}\right|
\dfrac{m_b}{m_s}
\left|
V_{cb} + \sqrt{\dfrac{m_c}{m_t}} e^{-i\theta_\alpha}
\right|
\label{relation2}\,.
\end{equation}
\noindent  These   are  satisfied up to $O(\l^2)$.

%%%%%%%%%%%%%%%%%%%%%%%%%%%%%%%%%%%%%%%%%%%%%%%%%%%%%%%%%%%%%%%%%%
%%%%%%%%%%%%%%%%%%%%%%%%%%%%%%%%%%%%%%%%%%%%%%%%%%%%%%%%%%%%%%%%%%
Furthermore, by using eqs.(\ref{Vus}), (\ref{Vcb}), (\ref{Vub})
 and (\ref{unitary_e}),
we obtain the third relation as follows:
\begin{eqnarray}
 \tan \theta_{\mu \tau}^e &=&
 \dfrac{m_b}{m_s}
\left|
\dfrac{V_{ub}}{V_{us}}
\right|
\left|
1-
\dfrac{1}{y z}
\sqrt{\dfrac{m_s}{m_d}}\sqrt{\dfrac{m_u}{m_c}}
\left(
e^{i\theta}
+
\dfrac{y^2}{\sqrt{1-y^4}}
\dfrac{m_b^{}}{m_s^{}}
V_{ub}
\right)
\right| 
 \nonumber \\
&=&
 \dfrac{m_b}{m_s}
\left|
\dfrac{V_{ub}}{V_{us}}
\right|
\left| 1-
\left(
\dfrac{m_s^{}}{m_b^{}} e^{i\theta}
+ V_{ub} \right)
\left(
\sqrt{\dfrac{m_c^{}}{m_u^{}}}V_{ub} - V_{cb}
\right)^{-1}
\right|\ .
\label{relation3}
\end{eqnarray}

%%%%%%%%%%%%%%%%%%%%%%%%%%%%%%%%%%%%%%%%%%%%%%%%%%%%%%%%%%%
Up to  $O(\lambda)$ 
we obtain the fourth relation  as to $U^{\rm e}_{\mu 1}$ as follows:
\begin{equation}
\tan \theta_{\mu \tau}^e\left|U^{\rm e}_{\mu 1}\right|
=
\left|V_{ub} \left(\dfrac{1}{ V_{us}^2} \dfrac{m_d}{m_s}\right)\right|\,.
\label{relation4}
\end{equation}
%%%%%%%%%%%%%%%%%%%%%%%%%%%%%%%%%%%%%%%%%%%%%%%%%%%%%%%%%%%
%%%%%%%%%%%%%%%%  TEST %%%%%%%%%%%%%%%%%%%%%%%%%%%%%%%%%%%%

The neutrino mass matrix is unknown in our framework.
Assuming that the neutrino mass matrix is close to the diagonal matrix,
the lepton flavor mixing matrix (MNS matrix) is approximately 
\begin{equation}
 U_{\rm MNS} \simeq (U^{\rm e})^\dagger \ . 
\end{equation}
 \noindent  Then, the two relations
in eqs.(\ref{relation2}) and (\ref{relation3}) 
are testable by putting  experimental values.
%%%%%%%%%%%%%%%%%%%%%%%%%%%%%%%%%%%%%%%%%%%%%%%%%%%%
The second  relation, eq. (\ref{relation2}), is different from the ones
in refs. \cite{Albright,BKY}, where the the contribution of the up-quark 
sector is negligible. 
In the right hand side of  eq. (\ref{relation2}), the contribution of
the up-quarks is comparable to $V_{cb}$. Then, the phase $\theta_\a$
is an important ingredient in this relation.
We show  the allowed region of $\theta_\a$ versus $\sqrt{m_c/m_t}$
for $\tan\theta_{\mu\tau}^e=0.9, \ 1, \ 1.1$ in Fig.1,
in which  $m_s/m_b=1/40$, $V_{cb}=0.034$, $V_{us}=0.22$ 
and  $\sqrt{m_c/m_t}=0.035\sim 0.060$ are taken
at the GUT scale \cite{FK}.  
This relation is satisfied by taking $\theta_\a=130^\circ \sim 230^\circ$.
%%%%%%%%%%%%%%%%%%%%%%%%%%%%%%%%%%%%%%%%%%%%%%%%%%%%%%%%%%%%% 

The third one, eq.(\ref{relation3}), is firstly examined in our work
%%%%%%%%%%%%%%%%%%%%%%%%%%%%%%%%%%%%%%%%
\footnote{In the case of neglecting correction terms, the relation was
tested in ref.\cite{BKY}.}.
%%%%%%%%%%%%%%%%%%%%%%%%%%%%%%%%%%%%%%%%%
  We have found that the correction terms reach up to  $50\%$ of the
leading term.  We show the allowed region of
$\theta$ versus $V_{ub}$ in Fig.2.
 The atmospheric neutrino data  
$\sin^2 2\th_{\rm atm} \geq 0.88$ is completely consistent with these
two  relations if  relevant values of the  phases are taken.
%%%%%%%%%%%%%%%%%%%%%%%%%%%%%%%%%%%%%%%%%%%%

The fourth one is not testable because there is no data of
 $U^{\rm e}_{\mu 1}$.
The interesting relation is the first one, which is the relation
among the lepton mixings.
Since we find $|U^{\rm e}_{e 3}|\simeq |U^{\rm e}_{e 2}|\sim \sqrt{m_1/m_2}$
as seen in  eq.(\ref{unitary_e}), these elements are small ones.
If the contribution of the mixing angle from  the neutrino mass matrix
is small, we must take the SMA-MSW solution for the solar neutrinos.
The relation $|U^{\rm e}_{e 3}|=\tan \theta_{\mu\tau}^e|U^{\rm e}_{e 2}|$
 should  be tested in the near future 
%%%%%%%%%%%%%%%%%%%%%%%%%%%%%%%%%%%%%%%%%%%%%
\footnote{This relation has been  also
discussed in the $SO(10)$ model of ref.{\cite{Albright}}.}.
%%%%%%%%%%%%%%%%%%%%%%%%%%%%%%%%%%%%%%%%%%%%%%%%%
For the present the CHOOZ bound \cite{CHOOZ}
$|{U_{e 3}}|<0.16$  is only  available.
The long baseline experiments are expected to observe the
$|{U_{e 3}}|$ element in the near future

%%%%%%%%%%%%%%%%%%%%%%%%%%%%%%%%%%%%%%%%%%%%%%%%%%%%%%%%%%%%%%%%
%%%%%%%%%%%  Neutrino Mixing Effects %%%%%%%%%%%%%%%%%%%%%%%%%%
%%%%%%%%%%%%%%%%%%%%%%%%%%%%%%%%%%%%%%%%%%%%%%%%%%%%%%%%%%%%%%%%

Until now, we have neglected the contribution from the neutrino sector.
If above relations turn to be  not satisfied by  experimental data
in the future, we should consider the effect of the neutrino sector.
For example, the LMA-MSW, LOW and VO solutions are inconsistent
with above relations.
Let us consider the effect of the neutrino mass matrix in general.
The MNS matrix is defined as
\begin{equation}
 V_{\rm MNS} = \left(U^{\rm e}\right)^{\dagger}U^{\nu}{\cal P}\ ,
\end{equation}
where $U^{\nu}$ is the unitary matrix which diagonarizes the neutrino
mass matrix, and  ${\cal P}$ is the Majorana phase matrix which we neglect
in our paper.
Hereafter we write the MNS matrix without  ${\cal P}$ as $U_{\rm MNS}$.
%%%%%%%%%%%%%%%%%%%%%%%
The $U^{\nu}$ is parametrized as
\begin{eqnarray}
&& U^{\nu} =
\left(
\begin{array}{ccc}
1 & 0 &0 \\
0 & e^{i\varphi_2} &0 \\
0 & 0 & e^{i\varphi_3}\\
\end{array}
\right)
\nonumber \\
&&
\times
\left(
\begin{array}{ccc}
C_{13}C_{12} & C_{13}S_{12}&S_{13}e^{i\varphi_1} \\
-C_{23}S_{12} - C_{12}S_{13}S_{23}e^{-i\varphi_1}
&
C_{12}C_{23} - S_{12}S_{13}S_{23}e^{-i\varphi_1}
& 
C_{13}S_{23}\\
-C_{12}C_{23}S_{13}e^{-i\varphi_1} + S_{12}S_{23}
&
-C_{23}S_{12}S_{13}e^{-i\varphi_1} - C_{12}S_{23}
&
C_{13} C_{23} 
\end{array}
\right)\,,
\label{unitary_nu}
\end{eqnarray}
where $S_{ij}$ $(C_{ij})$ is  $\sin \theta_{ij}$ $(\cos \theta_{ij})$
 ($i$ and $j$ is the family  index).  In this parametrization,
the relevant MNS mxings are given as
\begin{eqnarray}
U_{e2} &=& U^{\rm e}_{e1} S_{12}C_{13} e^{i\theta^\prime}
         + U^{\rm e}_{e2}\left(
               e^{i\varphi_2}C_{12}C_{23}
             - e^{i\left(\varphi_2-\varphi_1\right)}S_{12}S_{13}S_{23}
           \right) \nonumber \\
        & & \hspace{5ex} - U^{\rm e}_{e3}\left(
               e^{i\left(\varphi_3-\varphi_1\right)}S_{12}S_{13}C_{23}
             + e^{i\varphi_3}C_{12}S_{23}
           \right)
\,,
\label{Ue2}
%\nonumber
 \\
U_{e3} &=& U^{\rm e}_{e1} S_{13} e^{i\phi_1} 
         + U^{\rm e}_{e2} C_{13} S_{23} e^{i\varphi_2} 
         + U^{\rm e}_{e3} C_{13} C_{23} e^{i\varphi_3} 
\,,
\label{Ue3}% \nonumber
 \\
U_{\mu 3} &=& U^{\rm e}_{\mu 1} S_{13} e^{i\phi_1}
            + U^{\rm e}_{\mu 2} C_{13} S_{23} e^{i\varphi_2}
            + U^{\rm e}_{\mu 3} C_{13} C_{23} e^{i\varphi_3} \,,
\label{Um3}
\end{eqnarray}
where $\theta^\prime =\theta_\alpha+\theta_\beta$ and
$\phi_1=\varphi_1+\theta_\alpha+\theta_\beta$.

Due to $U^{\rm e}_{e1} \simeq 1$ and $U^{\rm e}_{e2} \simeq U^{\rm e}_{e3}$, 
 eq.(\ref{Ue3}) is rewritten  as 
\begin{equation}
U_{e3} \simeq S_{13} e^{i\phi_1} 
         + U^{\rm e}_{e2} C_{13}
          \left(   S_{23} e^{i\varphi_2} 
                +  C_{23} e^{i\varphi_3} \right)\,.
\label{Ue3xx}
\end{equation}
Since $U_{e3}$ is small from the CHOOZ experiment,  
 $S_{13}$ should be very small. Then, we obtain approximately
\begin{equation}
U_{\mu 3} \simeq U^{\rm e}_{\mu 3} 
\left(
  S_{23} e^{i\varphi_2} + C_{23} e^{i\varphi_3}
\right)\,,
\end{equation}
\noindent where we used  $U^{\rm e}_{\mu 2} \simeq U^{\rm e}_{\mu 3}$
by taking $y^4\simeq 1/2$.  We  require that $U_{\mu 3}$ is maximal
as suggested from  the experiments.
Since  $U^{\rm e}_{\mu 3}$ is already maximal, we get a condition
\begin{equation}
\left|
  S_{23} e^{i\varphi_2} + C_{23} e^{i\varphi_3}
\right|\simeq 1 \ ,
\end{equation}
\noindent which leads to
$\theta_{23}=0$ or $\varphi_3 - \varphi_2 = \dfrac{\pi}{2}$.
On the other hand,  for $U_{e2}$ we have
\begin{equation}
U_{e2} \simeq
           S_{12} e^{i\theta^\prime}
         + U^{\rm e}_{e3}C_{12}
               \left(
               C_{23}e^{i\varphi_2}
             - S_{23}e^{i\varphi_3}
           \right)\, .
\label{Ue2x}
\end{equation}
\noindent
Then, we get the modified relation between $U_{e3}$ and $U_{e2}$ 
from eqs.(\ref{Ue3xx}) and (\ref{Ue2x}) as follows:
\begin{equation}
 U_{e3}
 \simeq
\tan \theta_{\mu\tau}
\dfrac{C_{13}}{C_{12}}
\dfrac{\left(C_{23}e^{i\varphi_3} + S_{23}e^{i\varphi_2}\right)}
      {\left(C_{23}e^{i\varphi_2} - S_{23}e^{i\varphi_3}\right)}
\left(
U_{e2} - e^{i\theta^\prime} S_{13}
\right)
+e^{i\phi_1}S_{13}\ , 
\label{Ue3x}
\end{equation}
\noindent in which $S_{23}\simeq 0$ or 
 $\varphi_3= \varphi_2 + \dfrac{\pi}{2}$.
This relation is reduced to 
\begin{equation}
 U_{e3}
 \simeq
\tan \theta_{\mu\tau}
\dfrac{C_{13}}{C_{12}}
\left(
U_{e2} - e^{i\theta^\prime} S_{13}
\right)
e^{i\left(\varphi_3-\varphi_2\right)}
+e^{i\phi_1}S_{13}\ ,
\end{equation}
\noindent  for   $S_{23}=0(C_{23}=1)$, and 
\begin{equation}
 U_{e3}
 \simeq
i
\tan \theta_{\mu\tau}
\dfrac{C_{13}}{C_{12}}
\left(
U_{e2} - e^{i\theta^\prime} S_{13}
\right)
+e^{i\phi_1}S_{13}\ ,
\end{equation}
\noindent
for  $\varphi_3=\varphi_2+\dfrac{\pi}{2}$.
In each cases, the relation between  $U_{e2}$ and $U_{e3}$ is
independent of  $S_{23}(C_{23})$.

%%%%%%%%%%%%%%%%%%%%%%%%%%%%%%%%%%%%%%%%%%%%
%%%%%%%%%%%%%%%%%%%%%%%%%%%%%%%%%%%%%%%%%%%%
As seen in eq.(\ref{unitary_e}), we get
 $U^{\rm e}_{e2}\simeq U^{\rm e}_{e3}\simeq 0.05$ by putting
experimental values $m_e/m_\mu=0.07$ and $y^2=\sqrt{1/2}$. 
If the SMA-MSW solution,  $U_{e2}=0.02\sim 0.05$ \cite{BKS},
is true, $S_{12}$ is expected to be at most of order $O(\l^2)$
from eq.(\ref{Ue2x}). 
 On the other hand, $U_{e3}$ depends on $S_{13}$
 as seen in  eq.(\ref{Ue3x}).
As far as the strong cancellation between the first term and the second
one is not occured, we expect  $U_{e3}\simeq 0.05\sim 0.16$.

If the LMA-MSW, LOW or  VO solution is true, 
$S_{12}$ is expected to be around $1/\sqrt{2}$.  Then, the MNS mixing
$U_{e2}$ is dominated by $S_{12}$ in the neutrino sector.  $U_{e3}$ is
also expected to be $0.05\sim 0.16$ as well as the SMA-MSW solution.

\section{Discussions and Summary}

%%%%%%%%%%%%%%%%%%%%  Akhmedov  Branco Rebelo %%%%%%%%%%%
We have presented the generic relation between  $U_{e2}$ and $U_{e3}$.
Akhmedov, Branco and  Rebelo \cite{Ak} have also  presented model-independent
   relations of them
in the basis of the charged lepton mass matrix being  diagonalized.
They have assumed that there is no  fine-tuning between  parameters of the
mass matrix, and so have predicted  $U_{e3}$ for each solar neutrino
solutions.  Our general result is different from their one,  because
$U_{e3}$ could be $0.05\sim 0.16$ independent of the solar neutrino
solutions.  The important point is that the fine-tuning depends on the basis
of the mass matrices.  Even if the fine-tuning is occured among the
elements of the neutrino mass matrix in the 
basis of the charged lepton mass matrix being  diagonalized, 
there is the other basis without fine-tuning, for example, 
the lopsided basis of the charged lepton mass matrix.
Let us move to the diagonal basis of  the  charged lepton mass
matrix from the lopsided basis through the bi-unitary transformation.
Then, (1-2) and (1-3) entries of the  neutrino mass matrix in the new basis
are almost same due to the nearly maximal rotation in  
the second and the third family space 
even if the original neutrino mass matrix is the
hierarhical one without the fine-tuning among  each entry.
This transformed neutrino mass matrix looks like to have a
fine-tuning of the parameters.  However, this fine-tuning is not
accidental.  The important issue is that one should take a preferred basis,
 where there is no accidental fine-tuning, for  the
mass matrices.
Thus, the possibility  of $U_{e3}=0.05\sim 0.16$ is justified 
independent of the solar neutrino solutions. 

%%%%%%%%%%%%%%%%%%%%%%%%%%%%%%%%%%%%%%%%%%%%%%%%%%%%%%%
%%%%%%%%%%%%%%%%%%%% CP Violation %%%%%%%%%%%%%%%%%%%%%
%%%%%%%%%%%%%%%%%%%%%%%%%%%%%%%%%%%%%%%%%%%%%%%%%%%%%%%
It may be helpful to comment on the CP violation in the lepton sector.
As seen in eqs.(\ref{Ue2}-\ref{Um3}), the CP violating phases appear
when the contribution of the neutrino mass matrix is taken account.
Therefore, there is no relation between CP violations in the leptons
and the quarks.
In order to discuss the magnitude of the CP violation of the leptons,
 we estimate the quantity $J_{CP}$ \cite{J}.
Neglecting the higher order terms of  $S_{13}$ and
 $U^{\rm e}_{e3}$, Jarlskog parameter $J_{CP}$ is obtained as follows:
\begin{eqnarray}
J_{CP} &=& - \dfrac{1}{2}C_{12}S_{12}C_{13}^2 
\left\{
S_{13} \left( S_{23}^2\sin\left(\varphi_1 - \varphi_2 + \varphi_3\right)
         - C_{23}^2\sin\left(\varphi_1 + \varphi_2 - \varphi_3\right)
    \right) \right.\nonumber \\
& & \hspace{8ex}+
U^{\rm e}_{e3}C_{13}
\left(
  C_{23}^3        \sin\left(\theta^\prime - \varphi_2 \right) 
- C_{23}S_{23}^2  \sin\left(\theta^\prime + \varphi_2 - 2\varphi_3\right)
\right.
\nonumber \\
& & \hspace{18ex}
\left. \left.
- C_{23}^2S_{23}  \sin\left(\theta^\prime - 2\varphi_2+ \varphi_3 \right)
+ S_{23}^3        \sin\left(\theta^\prime - \varphi_3 \right)
\right)
\right\}\ ,
\end{eqnarray}
\noindent which is reduced to  
\begin{eqnarray}
J_{CP} &=& \dfrac{1}{2}C_{12}S_{12}C_{13}^2 
\left\{
       S_{13} \sin\left(\varphi_1 + \varphi_2 - \varphi_3\right)
    - U^{\rm e}_{e3}C_{13} \sin\left(\theta^\prime - \varphi_2 \right) 
\right\}\ ,
\end{eqnarray}
\noindent  for $S_{23}=0~(C_{23}=1)$, and
\begin{eqnarray}
J_{CP} &=& - \dfrac{1}{2}C_{12}S_{12}C_{13}^2 
\left\{
  S_{13}\cos{\varphi_1}
+ U^{\rm e}_{e3} C_{13} \sin\left(\theta^\prime - \varphi_2 - \theta_{23}
 \right)
\right\}\,,
\end{eqnarray}
\noindent for $\varphi_3=\varphi_2+\dfrac{\pi}{2}$.
Here  $\theta_{23}$ is mixing angle of the $S_{23}~(C_{23})$.
Thus, the CP violating effect depends on $S_{12}$ and $S_{13}$,
which comes from the neutrino mass matrix.

%%%%%%%%%%%%%%%%  Summary  %%%%%%%%%%%%%%%%%%%%%%%
%%%%%%%%%%%%%%%%%%%%%%%%%%%%%%%%%%%%%%%%%%%%%%%%%%
We have studied  implications of the generic
lopsided mass matrix of the charged leptons
by taking the SU(5) GUT relation in the NNI basis.
  We have found four  interesting relations among  the lepton mixings
and the quark ones, which are independent of the details of the model,
 neglecting effects of neutrino mass matrix.
Two relations among them are  examined by using the experimental data.
These are  satisfied by the experimental data.
   We have also discussed the relation between  $U_{e2}$ and $U_{e3}$
 incuding the contribution from the neutrino mass matrix.  
We have presented  the probable value
  $U_{e3}=0.05\sim 0.16$, which is  
independent of the solar neutrino solutions. 
The $U_{e3}$ is expected to  be measured in the long baseline 
experiments in the near future.

\vskip 1 cm
%%%%%%%%%%%%%%%%%%%%%%%%%%%%%%%%%%%%%%%%%%%%%%%%%%%%%%%%%%%%%%%%%%%%%%%%%%%%%
%%%%%%%%%%%%%%%%%%%%%%%%%%%%%%%%%%%%%%%%%%%%%%%%%%%%%%%%%%%%%%%%%%%%%%%%%%%%%
\section*{Acknowledgements}
 We would like to thank M. Bando for useful discussions. We also
give our thanks for kind hospitality at the post Summer Institute for
neutrino physics at Fuji-Yoshida in August, 2000. 
 This research is  supported by the Grant-in-Aid for Science Research,
  Ministry of Education, Science and Culture, Japan(No.10640274,
No.12047220).  
 The work of N.O. is supported by the JSPS Research Fellowship
for Young Scientists, No.2996.

\newpage
%%%%%%%%%%%%%%%%%%%%%%%%%%%%%%%%%%%%%%%%%%%%%%%%%%%%%%%%%%%%%%%%%%%%%%%%%%%%
%%%%%%%%%%%%%%%%%%%%%%%%%%%%%%%%%%%%%%%%%%%%%%%%%%%%%%%%%%%%%%%%%%%%%%%%%%%%

\newpage

%%%%%%%%%%%%%%%  Figure 1  %%%%%%%%%%%%%%%%%
\begin{figure}
\epsfxsize=12 cm
\centerline{\epsfbox{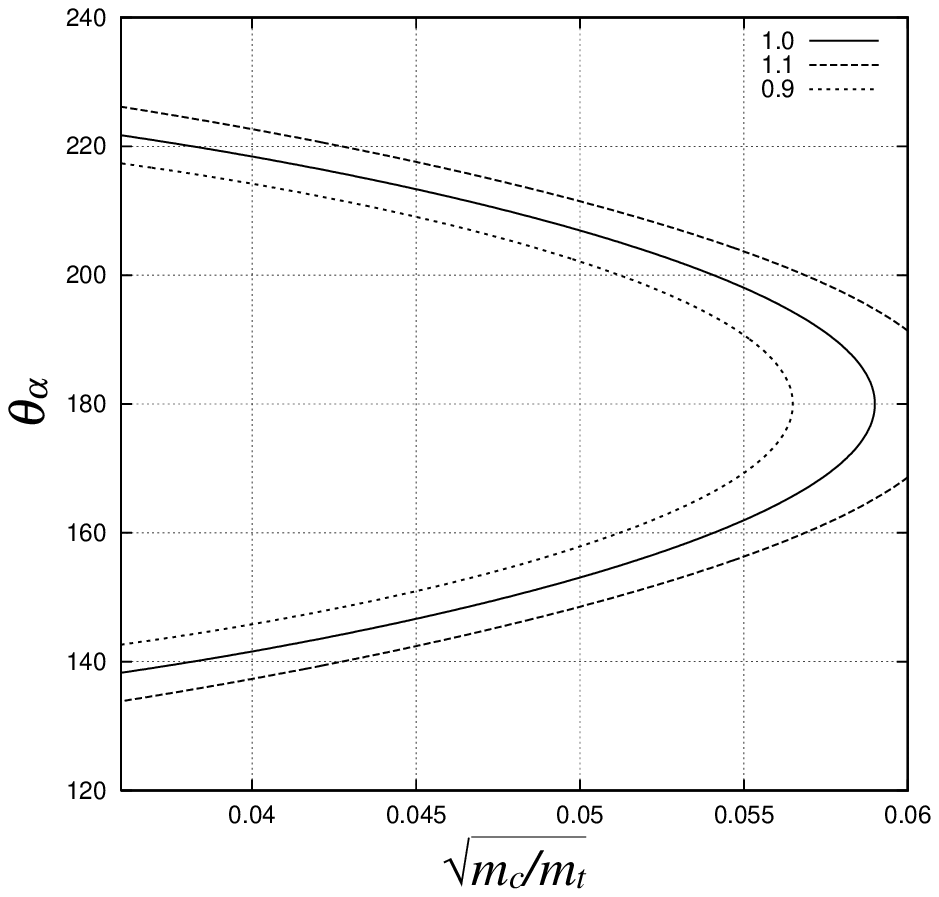}}
\caption{}
\end{figure}

Fig.1: The allowed curves  of $\theta_\a$ versus $\sqrt{m_c/m_t}$
for $\tan \theta_{\mu\tau}=0.9,\ 1.0,\ 1.1$
 in  the relation of eq.(\ref{relation2}). 
We fix  $m_s/m_b=1/40$, $V_{cb}=0.034$ and $V_{us}=0.22$ 
at the GUT scale.  

\newpage
%%%%%%%%%%%%%%%%%%%%%%%%%%%%%%%%%%%%%%%%%%%%% 
%%%%%%%%%%%%%%%  Figure 2  %%%%%%%%%%%%%%%%%
\begin{figure}
\epsfxsize=12 cm
\centerline{\epsfbox{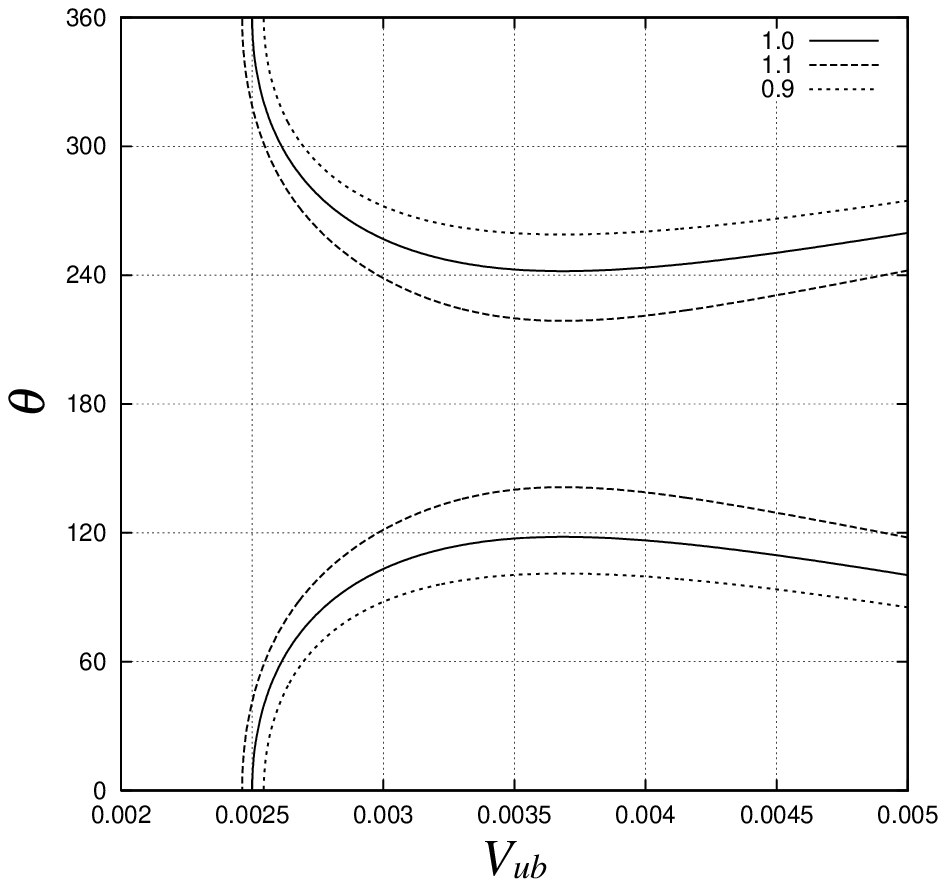}}
\caption{}
\end{figure}

Fig.2: The allowed curves of $\theta$ versus $V_{ub}$
for $\tan \theta_{\mu\tau}=0.9,\ 1.0,\ 1.1$
 in   the relation of  eq.(\ref{relation3}). 
 The  $m_s/m_b=1/40$, $V_{cb}=0.034$ and $V_{us}=0.22$ 
are  taken at the GUT scale.  

%%%%%%%%%%%%%%%%%%%%%%%%%%%%%%%%%%%%%%%%%%%%% 

\end{document}